\PassOptionsToPackage{dvipsnames}{xcolor}
\documentclass[twocolumn]{aastex7}

\usepackage{amsmath}
\usepackage{subfigure}
\usepackage{float}
\usepackage{gensymb}
\usepackage{hyperref}

\begin{document}

\title{Identifying and characterizing extragalactic circum-CBC exoplanets with future gravitational-wave detectors }

\author[0000-0001-7197-8899]{Avinash Tiwari}
\affiliation{Inter-University Centre for Astronomy and Astrophysics, Post Bag 4, Ganeshkhind, Pune - 411007, India}
\email{avinash.tiwari@iucaa.in}

\author[0000-0002-4103-0666]{Aditya Vijaykumar}
\affiliation{Canadian Institute for Theoretical Astrophysics, University of Toronto, 60 St. George Street, Toronto, ON M5S 3H8, Canada}
\email{avijaykumar@cita.utoronto.ca}

\author[0000-0001-5318-1253]{Shasvath J. Kapadia}
\affiliation{Inter-University Centre for Astronomy and Astrophysics, Post Bag 4, Ganeshkhind, Pune - 411007, India}
\email{shasvath.kapadia@iucaa.in}

\author[0000-0002-3680-2684]{Sourav Chatterjee}
\affiliation{Tata Institute of Fundamental Research, Homi Bhabha Road, Navy Nagar, Colaba, Mumbai 400005, India}
\email{sourav.c@tifr.res.in}

\begin{abstract}
Exoplanets are high-value targets for a variety of ground and space-based telescopes. All known exoplanets are Galactic, and a fraction of them orbit compact objects. 
In this work, we investigate the possibility of detecting extragalactic exoplanets orbiting stellar-mass compact binary coalescences (CBCs), such as binary neutron stars, neutron star-black holes, and binary black holes, using future gravitational wave (GW) detectors, including A+ (LIGO in O5), Einstein Telescope, and DECIGO.
We use the technique of reconstructing an external potential's profile by extracting information about the centre-of-mass (CoM) kinematics of a CBC encoded in the GWs it emits. In this work, the external potential is provided by the circum-CBC exoplanet, and the resulting signature on the GW waveform comes from the ``wobble'' of the CBC's CoM around the CBC-exoplanet barycentre. As a proof of principle, we consider a few example CBCs detectable with future detectors and a range of circum-CBC exoplanet parameters in circular and eccentric orbits. We find that for a significant fraction of the range of parameters considered, we can identify the presence of a circum-CBC exoplanet by extracting its mass (up to an unknown orbital inclination angle) within a factor $\mathcal{O}(1)$ of its true value, at $68\%$ confidence. 

\end{abstract}

\section{Introduction} \label{sec:intro}
To date, ${\sim}6000$ exoplanets have been discovered~\citep{ps}, most of which ($\sim 5520$) orbit around single stars, a small fraction ($\sim 480$) orbit around binaries, while a handful of them ($\sim 70$) orbit around trinaries. Though nearly all of these exoplanets orbit luminous, non-compact stars, a few have been found orbiting compact stars. In fact, the very first observation of an exoplanet involved two planets orbiting the millisecond pulsar PSR B1257+12 \citep{wolszczan1992pulsar}, which preceded the first detection of an exoplanet orbiting a main-sequence star (51 Pegasi b) \citep{mayor1995jupiter}. Detections of exoplanets around compact binaries, called circum-binary exoplanets, have also been claimed, including PSR B1620-26 b orbiting a Pulsar + White Dwarf binary \citep{thorsett1999pulsar,2003Sci...301..193S,  Fregeau_2006}, and NN Serpentis orbiting a White Dwarf + Red Dwarf binary \citep{beuermann2010nnser}.

All exoplanets have so far been detected using methods such as radial velocity \citep{mayor1995jupiter}, microlensing \citep{bond2004microlensing}, direct imaging \citep{marois2008direct}, transit timing variations \citep{holman2005ttv}, pulsar timing variations \citep{wolszczan1992pulsar}, and astrometry \citep{benedict2002astrometry}. A variety of electromagnetic telescopes and corresponding surveys are used for this purpose, including space-based missions such as Kepler \citep{borucki2010kepler}, TESS \citep{ricker2015tess}, CoRot \citep{cabrera2015corot}, JWST \citep{gardner2006jwst}, and ground-based surveys such as HARPS \citep{mayor2003harps}, HATNet and HATSouth \citep{bakos2018hatnet}, and SuperWASP \citep{hellier2017wasp}. Detecting {\it extragalactic} exoplanets using the methods and missions described above is out of reach at present. In fact, the most distant exoplanets known to date ($\sim 8.5$ kpc), SWEEPS-4 b and SWEEPS-11 b~\citep{2006Natur.443..534S}, are in the Milky Way.

Gravitational waves (GWs) could open up a new detection avenue on this front.
Indeed, it has been proposed that exoplanets orbiting Galactic binary white dwarfs (BWDs), an important source for Laser Interferometer Space Antenna (LISA)~\citep{Robson_2019}, may be detected using the imprint of the wobble of their centre-of-mass (CoM) on the GWs they emit \citep{lamberts2019predicting}. This is because the wobble of the BWD will imprint itself on the GW due to a time-varying Doppler shift, causing the observed GW frequency to drift. Exploiting this frequency drift at linear order in time, \cite{tamanini2018listening, tamanini2019gravitational, danielski2020will} argue that a combination of masses, degenerate with the orbital inclination, can be extracted from the GW signal. However, ascertaining the presence of an exoplanet will require subsequent follow-up with EM observations. 

In this work, we demonstrate that if an exoplanet orbits an \textit{extragalactic} compact binary coalescence (CBC), \textit{detailed} information about the planet can be extracted from the system's GW signal using future detectors.
The possibility of observing extragalactic exoplanets with GWs was briefly pointed out by \citep{seto2008detecting}, but it does not dwell on the possibility of extracting detailed information such as exoplanet mass and orbital radius / semi-major axis.

As a proof of concept, we consider ({\it i}) a BNS in the LIGO band at A+ sensitivity, ({\it ii}) a BNS and an NSBH in the ET band, and ({\it iii}) a couple of BBHs and a BNS in the DECIGO band. 
We consider a range of exoplanet masses, distances from the centre of the CBC, as well as circular and eccentric orbits. We find that, for a significant fraction of the assumed exoplanet parameter space, we can extract its mass and orbital radius with a fractional uncertainty of $\mathcal{O}(1)$.

\begin{figure}[ht!]
    \centering
    \includegraphics[width=0.995\linewidth]{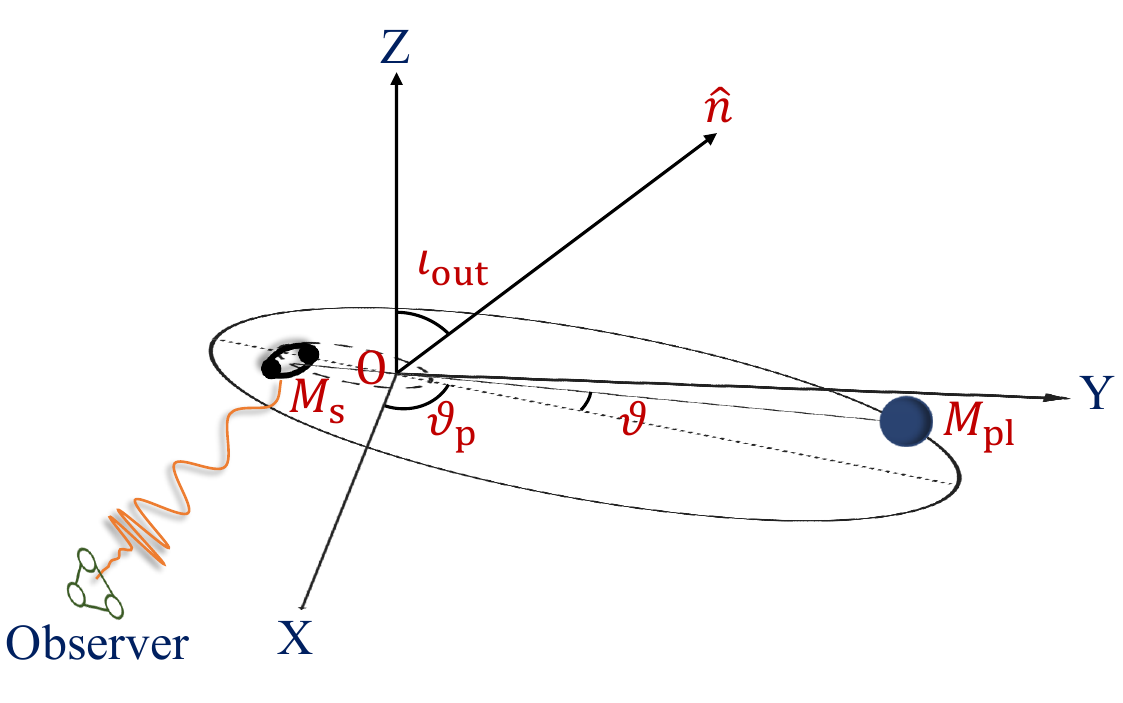}
    \caption{A schematic representation of an exoplanet (solid) and CBC (dashed) orbiting in eccentric orbits around the system's centre of mass (barycenter) $O$. $M_{\rm s}$ is the total mass of the CBC, $M_{\rm pl}$ is the mass of the exoplanet, $\vartheta_{\rm p}$ is the angular position of the periapsis from the X-axis (longitude of periapsis), $\vartheta$ is the angular position of the exoplanet relative to the periapsis (true anomaly), and $\iota_{\rm out}$ is the angle between angular momentum of the outer orbit and the observer's LOS $\Hat{n}$.}
    \label{fig: schem_exo}
\end{figure}

\section{Method}\label{sec: method}
Let $m_{1,\rm s}$ and $m_{2,\rm s}$ be the source frame masses of the primary
and secondary components of a CBC, respectively, $M_{\rm s} = m_{1,\rm s} + m_{2,\rm s}$ be the source frame total mass, $M_{\rm pl}$ be the mass of the exoplanet, $a$ be the semi-major axis of the exoplanet's orbit around the CBC, and $\Omega \equiv \sqrt{G (M_{\rm s} + M_{\rm pl}) / a^3 }$ and $e_{\rm out}$ be the mean motion and eccentricity, respectively, of the same. Then we can write the line-of-sight velocity (LOSV) of the centre of mass (CoM) of the CBC around the CBC-exoplanet \textit{barycentre} as (see Equation~(11) of the chapter \textit{Radial Velocity Techniques for Exoplanets} of \cite{Exoplanets_s_seager_book_2010}):
\begin{equation}
    \label{eq: losv_EO}
    V_{\rm L} = \frac{V_{\rm L,0}  \left[\cos \left(\vartheta+\vartheta_{\rm p}\right) + e_{\rm out} \cos \vartheta_{\rm p}\right]}{\sqrt{1-e_{\rm out}^2}}
\end{equation}
where $V_{\rm L,0} \equiv k a \Omega$, $ k \equiv M_{\rm pl} \sin \iota_{\rm out} / (M_{\rm s} + M_{\rm pl}) $, $\vartheta_{\rm p}$ is the longitude of periapsis\footnote{For the reasons mentioned in~\cite{Tiwari:losv2026}, we have set the longitude of the ascending node to 0.}, $\vartheta$ is the true anomaly, and $\iota_{\rm out}$ is the angle between our LOS and angular momentum of the outer orbit as depicted in Figure~\ref{fig: schem_exo}. The LOSV of the CBC's CoM in a circular outer orbit can be found by setting $e_{\rm out} = 0$, $\vartheta_{\rm p} = 0$, and $\vartheta \equiv \Omega_{\rm det} (t_{\rm u} - t_{\rm c}) + \theta_{\rm c}$ in Equation~\ref{eq: losv_EO}, where $t_{\rm u}$ is the time in the observer's frame accounting for only cosmological redshift, $t_{\rm c}$ is the time at coalescence, $\Omega_{\rm det} \equiv \Omega / (1 + z_{\rm cos})$\footnote{The extra factor $1 / (1 + z_{\rm cos})$ is due to cosmological time dilation.}, $\theta_{\rm c}$ is the location of the CBC in the outer orbit relative to the X-axis at the time of coalescence, and $z_{\rm cos}$ is the cosmological redshift. Specifically, it takes the form $V_{\rm L} = V_{\rm L,0} \cos (\Omega_{\rm det} (t_{\rm u} - t_{\rm c}) + \theta_{\rm c})$. Note that $\Omega$ becomes the angular frequency in this case, and $a$ becomes the exoplanet's orbital radius around the CBC.

\begin{figure*}[ht!]
    \centering
    \includegraphics[width=0.975\linewidth]{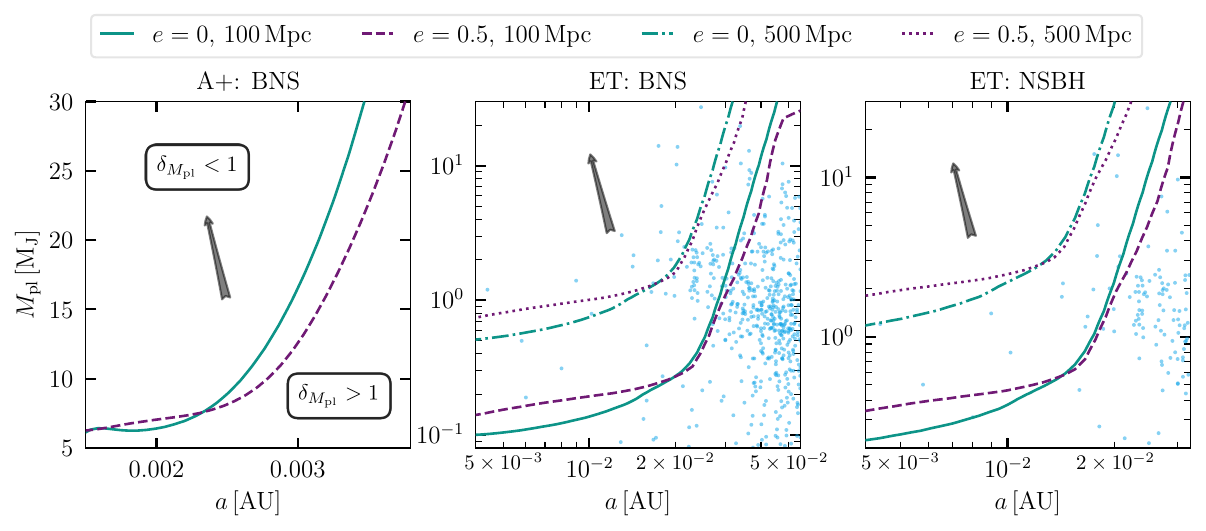}
    \caption{Detected exoplanets overplotted together with the $\delta_{M_{\rm pl}} = 1$ contours over a grid of $M_{\rm pl}$ and $a$ for the systems considered in A+ and ET band. The \textit{left} and {\it right panels} correspond to a 1.6-1.3 $M_{\odot}$ BNS in A+ and ET, respectively, while the \textit{right panel} corresponds to a 5-1.4 $M_{\odot}$ NSBH in ET. The solid contours correspond to the circular outer orbits of CBCs at 100 Mpc, the dashed ones correspond to eccentric outer orbits of CBCs at 100 Mpc, the dash-dotted ones correspond to circular outer orbits of CBCs at 500 Mpc,  while the dotted ones correspond to eccentric outer orbits of CBCs at 500 Mpc. The arrows represent the direction of increasing precision in the measurement of $M_{\rm pl}$, while the regions shown by $\delta_{M_{\rm pl}} > 1$, in the bottom of the contours, represent the parameter space where we cannot ascertain whether the third body is an exoplanet or a more massive object.}
    \label{fig: Exoplanet_overplot_plot}
\end{figure*}

Let $z_{\rm L} \equiv  V_{\rm L} / c$ be the Doppler shift due to the LOSV of the CBC's CoM and $z_{\rm L,0} \equiv V_{\rm L,0} / c$ be the corresponding maximum Doppler shift, where $c$ is the speed of light. Under the stationary phase approximation (SPA) and the assumption that $z_{\rm L, 0} \ll 1$ for circular outer orbits and $z_{\rm L, 0}/\sqrt{1 - e_{\rm  out}^2} \ll 1$ for eccentric outer orbits, the GW waveform of the CBC moving with a time-varying relative velocity can be written as:
\begin{equation}
    \label{eq:WF}
    \tilde{h}_{\rm TV}(f) = \tilde{h}(f) e^{i \Delta \Psi_{4} (f;z_{\rm L})}
\end{equation}
where $\tilde{h}(f)$ is the unmodulated GW waveform~\citep{PhysRevD.80.084043} and $\Delta \Psi_{4} (f; z_{\rm L})$ is the phase correction due to the LOSV, where 4 in the subscript indicates that the leading-order corrections appear at the 4th post-Newtonian order (4 PN).

For circular outer orbits, at the leading order, $\Delta \Psi_{4} (f; z_{\rm L})$ can be written as~\citep{Tiwari:losv2026}:  
\begin{multline}
    \label{eq: ph_cor_losv_circ}
    \Delta \Psi_{\rm 4,C} (f) = - \frac{5 z_{\rm L,0}}{128 \eta} \frac{v^3}{\xi} \Biggl[\sin \left(\frac{\xi}{v^8} - \theta_{\rm c} \right) \\ - \sin \left(\frac{\xi}{v_{\rm lso}^8} - \theta_{\rm c} \right)  \Biggr]
\end{multline}

where $v \equiv (\pi G M f / c^3)^{1/3}$, $v_{\rm lso} \equiv (\pi G M f_{\rm lso} / c^3)^{1/3}$, and $\xi \equiv (5 \Omega_{\rm det} / (256 \eta)) G M / c^3$. Here $m_1 = (1 + z_{\rm cos})m_{1,\rm s}$, $m_2 = (1 + z_{\rm cos})m_{2,\rm s}$, $M \equiv m_1 + m_2$ is the cosmologically redshifted total mass of the CBC, $\eta \equiv m_1m_2/M^2$ is the symmetric mass ratio of the CBC, $f$ is the observed GW frequency, and $f_{\rm lso}$ is the same at the last stable orbit. For eccentric outer orbits, $\cos \vartheta$ and $\sin \vartheta$ appearing in Equation~\ref{eq: losv_EO} can be expanded in eccentric harmonics in terms of the mean anomaly\footnote{Note that, in this case, $\theta_{\rm c}$ becomes the mean anomaly at the time of coalescence.} $\zeta \equiv \Omega_{\rm det} (t_{\rm u} - t_{\rm c}) + \theta_{\rm c}$. The phase correction in this case is given by Equation (20) of~\cite{Tiwari:losv2026}. Note that there will be amplitude corrections as well. However, as argued in~\cite{Tiwari:losv2026}, we will not be including the amplitude corrections in the Fisher matrix because these will be negligible.

\begin{figure*}[ht!]
    \centering
    \includegraphics[width=0.975\linewidth]{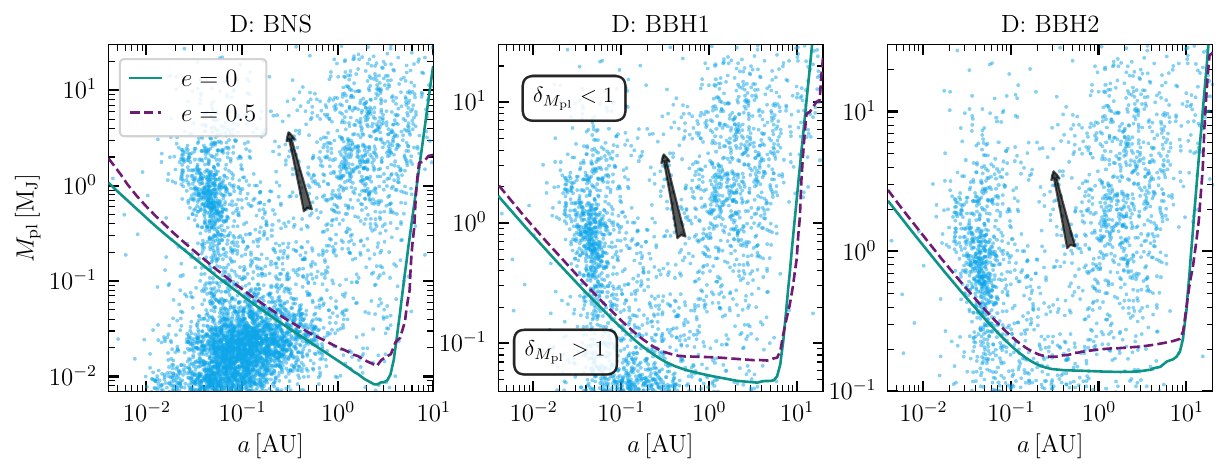}
    \caption{Detected exoplanets overplotted together with the $\delta_{M_{\rm pl}} = 1$ contours over a grid of $M_{\rm pl}$ and $a$ for the systems considered at 1 Gpc in the DECIGO band. The \textit{left panel} corresponds to a 1.6-1.3 $M_{\odot}$ BNS, the {\it middle panel} corresponds to a 10 - 9 $M_{\odot}$ BBH, while the \textit{right panel} corresponds to a 30-25 $M_{\odot}$. The solid contours correspond to the circular outer orbits, while the dashed ones correspond to eccentric outer orbits with $e_{\rm out} = 0.5$. The arrows have the same meaning as in Figure~\ref{fig: Exoplanet_overplot_plot}.}
    \label{fig: Exoplanet_overplot_plot_D}
\end{figure*}

To measure the mass of the exoplanet and its other orbital parameters, we perform the Fisher matrix analysis~\citep{CutlerFlanagan} on $\boldsymbol{\Theta}_{\rm E} \equiv \{\ln D_{\rm L},\, \ln \mathcal{M},\, \ln \eta,\, z_{\rm L,0},\, \ln \Omega_{\rm det},\, e_{\rm out},\, \theta_{\rm c},\, \vartheta_{\rm p}\}$ for eccentric outer orbits and $\boldsymbol{\Theta}_{\rm C} \equiv \{\ln D_{\rm L},\, \ln \mathcal{M},\, \ln \eta,\, z_{\rm L,0},\, \ln \Omega_{\rm det},\, \theta_{\rm c}\}$ for circular outer orbits.
Here, $D_{\rm L}$ and $\mathcal{M} \equiv M \eta^{3/5}$ are the luminosity distance and cosmologically redshifted chirp mass of the CBC, respectively. We then invert the Fisher matrices to acquire the corresponding covariance matrices $\boldsymbol{\Sigma}_{\rm E}$ and $\boldsymbol{\Sigma}_{\rm C}$, which contain the errors in the measurements of parameters $\boldsymbol{\Theta}_{\rm E}$ and $\boldsymbol{\Theta}_{\rm C}$, respectively. Finally, we transform $\boldsymbol{\Sigma}_{\rm C}$ to the covariance matrix in terms of $M_{\rm pl}$ and $a$ using $(\boldsymbol{J}^{-1}) (\boldsymbol{J}_{\ln}^{-1}) \boldsymbol{\Sigma}_{\rm C,S} (\boldsymbol{J}_{\ln}^{-1})^{\rm T} (\boldsymbol{J}^{-1})^{\rm T}$, where $\boldsymbol{\Sigma}_{\rm C,S}$ is the sub-matrix of $\boldsymbol{\Sigma}_{\rm C}$ corresponding to the rows $\{ \ln \mathcal{M},\, \ln \eta,\, z_{\rm L,0},\, \ln \Omega_{\rm det} \}$, $\boldsymbol{J}_{\ln} \equiv \partial (\ln \mathcal{M},\, \ln \eta,\, z_{\rm L,0},\, \ln \Omega_{\rm det}) / \partial (\mathcal{M},\, \eta,\, z_{\rm L,0},\, \Omega_{\rm det})$ is the Jacobian of the transformation from $(\mathcal{M},\, \eta,\, z_{\rm L,0},\, \Omega_{\rm det})$ to $(\ln \mathcal{M},\, \ln \eta,\, z_{\rm L,0},\, \ln \Omega_{\rm det} \})$, and $\boldsymbol{J} \equiv \partial(\mathcal{M},\, \eta,\, z_{\rm L,0},\, \Omega_{\rm det}) / \partial(\mathcal{M},\, \eta,\, M_{\rm pl},\, a)$ is the same of the transformation from $(\mathcal{M},\, \eta,\, M_{\rm pl},\, a)$ to $( \mathcal{M},\, \eta,\, z_{\rm L,0},\, \Omega_{\rm det})$. $\boldsymbol{J}_{\ln}^{-1}$ is simply ${\rm diag}(\mathcal{M},\, \eta,\, 1,\, \Omega_{\rm det})$, while $\boldsymbol{J}$ and $\boldsymbol{J}^{-1}$ are given by Equations~\ref{eq: jac} and~\ref{eq: jac_inv} of the Appendix~\ref{app: jac}, respectively. We follow the same procedure for eccentric outer orbit cases as well, with the $ \boldsymbol{\Sigma}_{\rm C,S}$ replaced by $ \boldsymbol{\Sigma}_{\rm E,S}$ corresponding to the same variable as in the circular outer orbit case. 

An alternate way to get the constraints on $M_{\rm pl}$ and other outer orbit parameters is to directly sample these parameters through the log-likelihood of $\boldsymbol{\Theta}_{\rm C}$ and $\boldsymbol{\Theta}_{\rm E}$, which is provided by the covariance matrix because it gives the Gaussian approximation to the GW likelihood. We use \textsc{emcee}~\citep{emcee}, and set uniform priors on all parameters similar to \cite{Tiwari:2024pvb}. We refer the reader to Section IIC of \cite{Tiwari:2024pvb} for more details of the method. We show three examples of this method as well in the Results section. Throughout this {\it Letter}, we assume that $\sin \iota_{\rm out} = 1$ and is known. Note that $\sin \iota_{\rm out}$ is degenerate with $M_{\rm pl}$ and $a$, and this degeneracy cannot be broken. Therefore, the errors in the relevant parameters should be treated as lower limits.

\begin{table*}[ht!]
    \centering
    \begin{tabular}{|c|c|c|c|c|c|c|c|}
        \hline
        \textbf{System} & \textbf{Exoplanet name} & $M_{\rm pl}\, [\rm M_J]$ & $a\,[\rm AU]$ & $e_{\rm out}$ & $t_{\rm obs}$ & $[f_{\rm min}, \, f_{\rm max}]$ & $\rho$ \\
        \hline
        \hline
        ET: BNS & \textit{Kepler-80 f} \citep{2016ApJ...822...86M} & 14.03188 & 0.0175 & 0.186 & 18.7 hrs & [2, 1483.4] & $\sim 758$ \\
        ET: NSBH & \textit{PSR J1719-1438 b} \citep{2011Sci...333.1717B} & 1.2 & 0.0044 & 0.06 & 7.24 hrs & [2, 672.1] & $\sim 1217$ \\
        D: BBH2 & \textit{GJ 676 A c} \citep{2016AandA...595A..77S} & 13.492 & 9.726 & 0.295 & 4 yr & [0.017, 10] & $\sim 42541$ \\
        \hline
    \end{tabular}
    \caption{Table of known galactic exoplanets considered in this analysis as examples. We place these exoplanets in fiducial orbits around CBCs in different detector bands. The tabulated values of $M_{\rm pl}$, $a$, and $e_{\rm out}$ are the ones inferred from EM observations, while the observation time and cut-off frequencies are chosen following section~\ref{sec: results}, and the optimal signal-to-noise ratios (SNRs) $\rho$ have been calculated following~\cite{Tiwari:2024pvb} assuming a face-on CBC at 100 Mpc for ET systems and 1 Gpc for the DECIGO system.}
    \label{tab: tab_det_exo_sys}
\end{table*}

\section{Results}\label{sec: results}
We consider exoplanets around: 
\begin{itemize}
    \item a $1.6 - 1.3 \,  M_{\odot}$ BNS observed in A+ (A+: BNS) at 100 Mpc,
    \item a $1.6 - 1.3 \,  M_{\odot}$ BNS observed in ET (ET: BNS) at 100 Mpc and 500 Mpc,
    \item a $5-1.4\,M_{\odot}$ NSBH observed in ET (ET: NSBH) at 100 Mpc and 500 Mpc,
    \item a $1.6 - 1.3 \,  M_{\odot}$ BNS observed in DECIGO (D: BNS) at 1 Gpc,
    \item a $10 - 9 \,  M_{\odot}$ BBH observed in DECIGO (D: BBH1) at 1 Gpc, and
    \item a $30 - 25 \,  M_{\odot}$ BBH observed in DECIGO (D: BBH2) at 1 Gpc.
\end{itemize}
We consider circular as well as eccentric outer orbit scenarios in all these systems and detector configurations. For circular as well as eccentric outer orbits, we fix $\theta_{\rm c} = 0.1 \, \rm rad$ in all system and detector configurations, while setting $e_{\rm out} = 0.5$ and $\vartheta_{\rm p} = 0.1 \, \rm rad$ in the eccentric outer orbit cases for the same. 

While performing the Fisher matrix analysis, we set the frequency ranges to 5 Hz - $f_{\rm lso}$ for all system and outer orbit configurations in A+, while we set the same in the case of ET to 2 Hz - $f_{\rm lso}$. For all system and outer orbit configurations in DECIGO, we choose a maximum observation time of $t_{\rm obs} = 4 \, \rm years$ and, accordingly, choose the maximum frequency $f_{\rm max}$ following~\cite{PhysRevD.71.084025} and minimum frequency $f_{\rm min}$ following the Section IIIB of~\cite{Tiwari:losv2026} to ensure the validity of SPA, assuming a $[10^{-2}, \,10]\, {\rm Hz}$ sensitivity band. We ensure that $z_{\rm exo,0} \equiv k_{\rm exo} a \Omega \leq 0.05$ to ensure $z_{\rm exo,0} \ll 1$ for circular outer orbits, while $z_{\rm exo,0} \equiv k_{\rm exo} a \Omega / \sqrt{1 - e_{\rm out}^2} \leq 0.05$ to ensure $z_{\rm exo,0}  / \sqrt{1 - e_{\rm out}^2} \ll 1$ for eccentric outer orbits. Here $ k_{\rm exo} \equiv M_{\rm s} \sin \iota_{\rm out} / (M_{\rm s} + M_{\rm pl})$. This choice ensures that the maximum Doppler shift due to the motion of the CBC's CoM is $\ll 1$. In addition, we also ensure that the CBC-exoplanet system is stable against the escape of the exoplanet. Specifically, we have used Equation (2) of~\cite{2022MNRAS.516.4146V}, which is Equation (90) of~\cite{2001MNRAS.321..398M}, with the mutual inclination of the inner and outer orbits $\iota_{\rm mut}$ to be $\pi / 2$ because we are assuming the inner orbit to be face-on, while the outer orbit to be edge-on. As mentioned earlier, this implies that the errors estimated in this work are lower limits: the edge-on outer orbit enhances the Doppler effect, while the face-on inner orbit of the CBC maximizes the SNR measured by the detectors. Note that we have further checked that effects like gravitational redshift and Shapiro delay are suppressed in comparison to the Doppler shift (see~\cite{Tiwari:losv2026} for more details).

\begin{figure}
    \centering
    \includegraphics[width=0.975\linewidth]{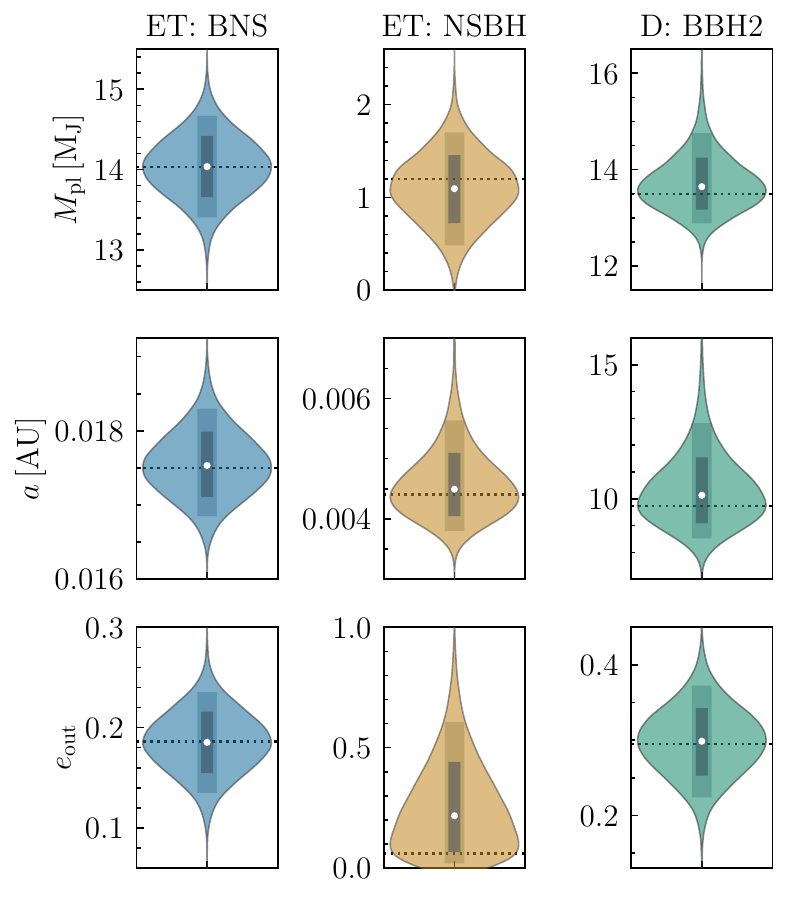}
    \caption{\textbf{$1d$ marginalised posteriors} of $M_{\rm pl}$, $a$, and $e_{\rm out}$ for example systems considered in eccentric orbits around a BNS in ET band (ET: BNS), NSBH in ET band (ET: NSBH), and BBH in DECIGO band (D: BBH2). The dashed lines represent the true values of the parameters. The shaded boxes within the violins represent the 68\% (smaller boxes) and 90\% (larger boxes) credible intervals, while the circles within the same represent the median values.}
    \label{fig: 1d_posts}
\end{figure}

Figure~\ref{fig: Exoplanet_overplot_plot} shows the known exoplanets\footnote{We use the catalogue of Exoplanets available at \href{https://exoplanetarchive.ipac.caltech.edu/cgi-bin/TblView/nph-tblView?app=ExoTbls\&config=PSCompPars}{exoplanetarchive.ipac}.} plotted over a grid of $M_{\rm pl}$ and $a$ for A+: BNS, ET: BNS, and ET: NSBH cases together with the $\delta_{M_{\rm pl}} \equiv \Delta M_{\rm pl} / M_{\rm pl} = 1$ contours, while Figure~\ref{fig: Exoplanet_overplot_plot_D} shows the same for D: BNS, D: BBH1, and D: BBH2 cases. We find that:
\begin{enumerate}
    \item none of the known exoplanets will be detected with A+, if they're orbiting a BNS at $100 \, {\rm Mpc}$.
    \item a significant number of the known exoplanets will be detected with ET, if they're orbiting a BNS at $100 \, {\rm Mpc}$, while a smaller fraction of them would be detected if they were orbiting an NSBH at the same $D_{\rm L}$. These numbers drop to a smaller number in the ET: BNS case, and to very few in the ET: NSBH case, when we increase $D_{\rm L}$ to 500 Mpc, as a direct consequence of the drop in SNR.
    \item a very large number of these exoplanets will be detected with DECIGO, if they're orbiting around a BNS or a BBH at $1 \, {\rm Gpc}$. 
\end{enumerate}
We further find a shift in the mass range of detectable exoplanets orbiting around the CBCs as we increase the CBC masses in all cases. This is because for more massive CBCs, only very massive exoplanets will be able to produce periodic LOSVs that are measurable. 

In Figure~\ref{fig: Exoplanet_overplot_plot_D}, we observe that the $\delta_{M_{\rm pl}}$ contours first fall sharply as a function of $a$, reach a minimum, and then rise sharply. The rise towards large $a$ can also be seen in Figure~\ref{fig: Exoplanet_overplot_plot}. This is because of the transition from the $\Omega_{\rm det} t_{\rm obs} / 2 \pi \gg 1$ region of the parameter space to the $\Omega_{\rm det} t_{\rm obs} / 2 \pi \ll 1$ region. Specifically, from Equation (\ref{eq: ph_cor_losv_circ}), we see that in the $\Omega_{\rm det} t_{\rm obs} / 2 \pi \gg 1$ region, the magnitude of the phase correction varies as $z_{\rm L,0} / \Omega_{\rm det} \propto a$ for a given $M_{\rm pl}$ because $z_{\rm L,0} \propto a^{-1/2}$ and $\Omega_{\rm det} \propto a^{-3/2}$ and hence, contrary to our expectation, we find that a low mass exoplanet which is closer to the CBC is not detectable, becomes detectable at a larger distance. As we keep increasing the distance, we hit the $\Omega_{\rm det} t_{\rm obs} / 2 \pi \sim 1$ region and enter the $\Omega_{\rm det} t_{\rm obs} / 2 \pi \ll 1$ region where the magnitude of the phase correction varies as $z_{\rm L,0} \Omega_{\rm det} \propto 1/a^2$ at the lowest order and therefore the same exoplanet again becomes non-detectable. At the same time, for a fixed $a$, the magnitude of the phase correction varies $\propto M_{\rm pl}$, assuming that $M_{\rm pl} + M_{\rm s} \approx M_{\rm s}$ is constant, in the whole parameter space, making larger masses more easily detectable. In combination, in the $\Omega_{\rm det} t_{\rm obs} / 2 \pi \gg 1$ region, the detectability of the exoplanet is determined by the quantity $M_{\rm pl}a$, while in the $\Omega_{\rm det} t_{\rm obs} / 2 \pi \ll 1$ region, the same is determined by $M_{\rm pl} / a^2$, which leads to the shapes of the contours in the Figures~\ref{fig: Exoplanet_overplot_plot} and~\ref{fig: Exoplanet_overplot_plot_D}.

For illustrative purposes, following the Figures~\ref{fig: Exoplanet_overplot_plot} and~\ref{fig: Exoplanet_overplot_plot_D}, we also consider three known exoplanets in fiducial eccentric orbits around CBCs in ET: BNS, ET: NSBH, and D: BBH2 configurations considered in this analysis. Table~\ref{tab: tab_det_exo_sys} contains the information about the known galactic exoplanets considered in this analysis as examples. Figure~\ref{fig: 1d_posts} shows the marginal posterior of $M_{\rm pl}$, $a$, and $e_{\rm out}$ for these systems. We find that the true values of all parameters are recovered at 90\% credible levels, except that 0 eccentricity can still not be ruled out in the ET: NSBH case. In Figures~\ref{fig: ET_BNS_corn_plot} and~\ref{fig: phase_corr_et_bns} of Appendix~\ref{app: add_fig}, we show an additional corner plot and frequency and time domain phase corrections for the ET: BNS example shown in Figure~\ref{fig: 1d_posts}. Note that for these examples, we have used the sampling method discussed towards the end of the Methods section using the priors mentioned in Table~\ref{tab: priors} in Appendix~\ref{app: priors}.

\section{Discussion}
Using the technique of reconstructing the profile of an external potential developed in~\cite{Tiwari:losv2026}, we have explored the possibility of detecting and characterizing extra-galactic exoplanets orbiting around BNSs and BBHs in future ground and space-based GW detectors: A+, ET, and DECIGO. We have demonstrated that a significant number of known exoplanets will be detectable by ET if they're orbiting around a BNS at $100 \, {\rm Mpc}$, and a larger number of them will be detectable by DECIGO if they're orbiting a BNS or a typical stellar-mass BBH at $1 \, {\rm Gpc}$. By placing three of the known Galactic exoplanets: \textit{Kepler-80 f}, \textit{PSR J1719-1438 b}, and \textit{GJ 676 A c} in fiducial orbits around a BNS detectable with ET, an NSBH detectable with ET, and a BBH detectable with DECIGO, respectively, we have further shown that we will be able to extract the masses of these exoplanets together with the sizes of the outer orbits.   

In comparison to EM observations, this method does not suffer from effects like dust extinction and scattering in the interstellar/intergalactic medium (ISM/IGM), and hence has the advantage of being sensitive to much larger distances, making it suitable for extragalactic exoplanet detection. While this method is mostly sensitive to the\footnote{See~\cite{Exoplanets_s_seager_book_2010} for more details on the Exoplanet subcategories.} \textit{Hot Neptunes} ($ 10 \, M_{\oplus} \lesssim M_{\rm pl} \lesssim 25 \, M_{\oplus}$),  \textit{Hot Saturns} ($ 25 \, M_{\oplus} \lesssim M_{\rm pl} \lesssim 150 \, M_{\oplus}$), and \textit{Hot Jupiters} ($M_{\rm pl} \gtrsim 150 \, M_{\oplus} \, \sim 0.5 \, {\rm M_J}$) orbiting the CBCs detectable with future detectors, \textit{Hot super-Earths} ($ 1 \, M_{\oplus} \lesssim M_{\rm pl} \lesssim 10 \, M_{\oplus}$ and within $\lesssim 0.1 \, \rm AU$ from the CBC) and \textit{super-Earths} ($ 1 \, M_{\oplus} \lesssim M_{\rm pl} \lesssim 10 \, M_{\oplus}$ and far away $\gtrsim 0.1 \, \rm AU$ from the CBC) orbiting BNSs can also be detected with the detectors like DECIGO. 

On the other hand, previous works on GW-based exoplanet detection were restricted to Galactic systems ~\citep{tamanini2018listening, tamanini2019gravitational, danielski2020will}, and could not extract detailed information about the exoplanet (such as its mass and the radius of its orbit). Our work therefore presents a step change in the ability to detect, identify, and characterize extragalactic exoplanets. If these are observed with future GW detectors, they would have far-reaching implications for the astrophysics of exoplanets and the formation of planetary systems in general.

\section*{Acknowledgements}
We thank Lalit Pathak and Nathan Johnson-McDaniel for useful discussions and suggestions. SJK acknowledges support from ANRF/SERB Grants SRG/2023/000419 and MTR/2023/000086.
\vspace{5mm}

\textit{Software}: \texttt{NumPy} \citep{vanderWalt:2011bqk}, \texttt{SciPy} \citep{Virtanen:2019joe}, \texttt{astropy} \citep{2013A&A...558A..33A, 2018AJ....156..123A}, \texttt{Matplotlib} \citep{Hunter:2007}, \texttt{jupyter} \citep{jupyter}.

\bibliographystyle{aasjournalv7}
\bibliography{references}

\begin{thebibliography}{}
\expandafter\ifx\csname natexlab\endcsname\relax\def\natexlab#1{#1}\fi
\providecommand{\url}[1]{\href{#1}{#1}}
\providecommand{\dodoi}[1]{doi:~\href{http://doi.org/#1}{\nolinkurl{#1}}}
\providecommand{\doeprint}[1]{\href{http://ascl.net/#1}{\nolinkurl{http://ascl.net/#1}}}
\providecommand{\doarXiv}[1]{\href{https://arxiv.org/abs/#1}{\nolinkurl{https://arxiv.org/abs/#1}}}

\bibitem[{ {Astropy Collaboration} {et~al.}(2013){Astropy Collaboration},
  {Robitaille}, {Tollerud}, {Greenfield}, {Droettboom}, {Bray}, {Aldcroft},
  {Davis}, {Ginsburg}, {Price-Whelan}, {Kerzendorf}, {Conley}, {Crighton},
  {Barbary}, {Muna}, {Ferguson}, {Grollier}, {Parikh}, {Nair}, {Unther},
  {Deil}, {Woillez}, {Conseil}, {Kramer}, {Turner}, {Singer}, {Fox}, {Weaver},
  {Zabalza}, {Edwards}, {Azalee Bostroem}, {Burke}, {Casey}, {Crawford},
  {Dencheva}, {Ely}, {Jenness}, {Labrie}, {Lim}, {Pierfederici}, {Pontzen},
  {Ptak}, {Refsdal}, {Servillat}, \& {Streicher}}]{2013A&A...558A..33A}
{Astropy Collaboration}, {Robitaille}, T.~P., {Tollerud}, E.~J., {et~al.} 2013,
  \bibinfo{title}{{Astropy: A community Python package for astronomy},} \aap,
  558, A33, \dodoi{10.1051/0004-6361/201322068}

\bibitem[{ {Astropy Collaboration} {et~al.}(2018){Astropy Collaboration},
  {Price-Whelan}, {Sip{\H{o}}cz}, {G{\"u}nther}, {Lim}, {Crawford}, {Conseil},
  {Shupe}, {Craig}, {Dencheva}, {Ginsburg}, {VanderPlas}, {Bradley},
  {P{\'e}rez-Su{\'a}rez}, {de Val-Borro}, {Aldcroft}, {Cruz}, {Robitaille},
  {Tollerud}, {Ardelean}, {Babej}, {Bach}, {Bachetti}, {Bakanov}, {Bamford},
  {Barentsen}, {Barmby}, {Baumbach}, {Berry}, {Biscani}, {Boquien}, {Bostroem},
  {Bouma}, {Brammer}, {Bray}, {Breytenbach}, {Buddelmeijer}, {Burke},
  {Calderone}, {Cano Rodr{\'\i}guez}, {Cara}, {Cardoso}, {Cheedella}, {Copin},
  {Corrales}, {Crichton}, {D'Avella}, {Deil}, {Depagne}, {Dietrich}, {Donath},
  {Droettboom}, {Earl}, {Erben}, {Fabbro}, {Ferreira}, {Finethy}, {Fox},
  {Garrison}, {Gibbons}, {Goldstein}, {Gommers}, {Greco}, {Greenfield},
  {Groener}, {Grollier}, {Hagen}, {Hirst}, {Homeier}, {Horton}, {Hosseinzadeh},
  {Hu}, {Hunkeler}, {Ivezi{\'c}}, {Jain}, {Jenness}, {Kanarek}, {Kendrew},
  {Kern}, {Kerzendorf}, {Khvalko}, {King}, {Kirkby}, {Kulkarni}, {Kumar},
  {Lee}, {Lenz}, {Littlefair}, {Ma}, {Macleod}, {Mastropietro}, {McCully},
  {Montagnac}, {Morris}, {Mueller}, {Mumford}, {Muna}, {Murphy}, {Nelson},
  {Nguyen}, {Ninan}, {N{\"o}the}, {Ogaz}, {Oh}, {Parejko}, {Parley}, {Pascual},
  {Patil}, {Patil}, {Plunkett}, {Prochaska}, {Rastogi}, {Reddy Janga},
  {Sabater}, {Sakurikar}, {Seifert}, {Sherbert}, {Sherwood-Taylor}, {Shih},
  {Sick}, {Silbiger}, {Singanamalla}, {Singer}, {Sladen}, {Sooley},
  {Sornarajah}, {Streicher}, {Teuben}, {Thomas}, {Tremblay}, {Turner},
  {Terr{\'o}n}, {van Kerkwijk}, {de la Vega}, {Watkins}, {Weaver}, {Whitmore},
  {Woillez}, {Zabalza}, \& {Astropy Contributors}}]{2018AJ....156..123A}
{Astropy Collaboration}, {Price-Whelan}, A.~M., {Sip{\H{o}}cz}, B.~M., {et~al.}
  2018, \bibinfo{title}{{The Astropy Project: Building an Open-science Project
  and Status of the v2.0 Core Package},} \aj, 156, 123,
  \dodoi{10.3847/1538-3881/aabc4f}

\bibitem[{M. {Bailes} {et~al.}(2011){Bailes}, {Bates}, {Bhalerao}, {Bhat},
  {Burgay}, {Burke-Spolaor}, {D'Amico}, {Johnston}, {Keith}, {Kramer},
  {Kulkarni}, {Levin}, {Lyne}, {Milia}, {Possenti}, {Spitler}, {Stappers}, \&
  {van Straten}}]{2011Sci...333.1717B}
{Bailes}, M., {Bates}, S.~D., {Bhalerao}, V., {et~al.} 2011,
  \bibinfo{title}{{Transformation of a Star into a Planet in a Millisecond
  Pulsar Binary},} Science, 333, 1717, \dodoi{10.1126/science.1208890}

\bibitem[{G.~{\'A}. Bakos {et~al.}(2018)Bakos {et~al.}}]{bakos2018hatnet}
Bakos, G.~{\'A}., {et~al.} 2018, \bibinfo{title}{The HATNet and HATSouth
  Exoplanet Surveys,} Handbook of Exoplanets, 1,
  \dodoi{10.1007/978-3-319-55333-7_111}

\bibitem[{G. Benedict {et~al.}(2002)Benedict {et~al.}}]{benedict2002astrometry}
Benedict, G., {et~al.} 2002, \bibinfo{title}{A mass for the extrasolar planet
  Gliese 876b determined from Hubble Space Telescope Fine Guidance Sensor 3
  astrometry and high-precision radial velocities,} The Astrophysical Journal,
  581, L115, \dodoi{10.1086/345794}

\bibitem[{E. Berti {et~al.}(2005)Berti, Buonanno, \& Will}]{PhysRevD.71.084025}
Berti, E., Buonanno, A., \& Will, C.~M. 2005, \bibinfo{title}{Estimating
  spinning binary parameters and testing alternative theories of gravity with
  LISA,} Phys. Rev. D, 71, 084025, \dodoi{10.1103/PhysRevD.71.084025}

\bibitem[{K. Beuermann {et~al.}(2010)Beuermann, Hessman, Dreizler, Marsh,
  Parsons, Winget, Miller, Schreiber, Kley, Dhillon, Littlefair, Copperwheat,
  \& Hermes}]{beuermann2010nnser}
Beuermann, K., Hessman, F.~V., Dreizler, S., {et~al.} 2010, \bibinfo{title}{Two
  planets orbiting the recently formed post-common envelope binary NN
  Serpentis,} Astronomy \& Astrophysics, 521, L60,
  \dodoi{10.1051/0004-6361/201015728}

\bibitem[{I. Bond {et~al.}(2004)Bond {et~al.}}]{bond2004microlensing}
Bond, I., {et~al.} 2004, \bibinfo{title}{OGLE 2003-BLG-235/MOA 2003-BLG-53: A
  planetary microlensing event,} The Astrophysical Journal, 606, L155,
  \dodoi{10.1086/421087}

\bibitem[{W.~J. Borucki {et~al.}(2010)Borucki {et~al.}}]{borucki2010kepler}
Borucki, W.~J., {et~al.} 2010, \bibinfo{title}{Kepler Planet-Detection Mission:
  Introduction and First Results,} Science, 327, 977,
  \dodoi{10.1126/science.1185402}

\bibitem[{A. Buonanno {et~al.}(2009)Buonanno, Iyer, Ochsner, Pan, \&
  Sathyaprakash}]{PhysRevD.80.084043}
Buonanno, A., Iyer, B.~R., Ochsner, E., Pan, Y., \& Sathyaprakash, B.~S. 2009,
  \bibinfo{title}{Comparison of post-Newtonian templates for compact binary
  inspiral signals in gravitational-wave detectors,} Phys. Rev. D, 80, 084043,
  \dodoi{10.1103/PhysRevD.80.084043}

\bibitem[{J. Cabrera {et~al.}(2015)Cabrera {et~al.}}]{cabrera2015corot}
Cabrera, J., {et~al.} 2015, \bibinfo{title}{Transiting exoplanets from the
  CoRoT space mission XXVII. CoRoT-28b, a planet orbiting an evolved star, and
  CoRoT-29b, a planet showing an asymmetric transit,} Astronomy \&
  Astrophysics, 579, A36, \dodoi{10.1051/0004-6361/201525580}

\bibitem[{C. Cutler \& E.~E. Flanagan(1994)Cutler \& Flanagan}]{CutlerFlanagan}
Cutler, C., \& Flanagan, E.~E. 1994, \bibinfo{title}{Gravitational waves from
  merging compact binaries: How accurately can one extract the binary's
  parameters from the inspiral waveform?} Phys. Rev. D, 49, 2658,
  \dodoi{10.1103/PhysRevD.49.2658}

\bibitem[{C. Danielski \& N. Tamanini(2020)Danielski \&
  Tamanini}]{danielski2020will}
Danielski, C., \& Tamanini, N. 2020, \bibinfo{title}{Will gravitational waves
  discover the first extra-galactic planetary system?} International Journal of
  Modern Physics D, 29, 2043007

\bibitem[{D. {Foreman-Mackey} {et~al.}(2013){Foreman-Mackey}, {Hogg}, {Lang},
  \& {Goodman}}]{emcee}
{Foreman-Mackey}, D., {Hogg}, D.~W., {Lang}, D., \& {Goodman}, J. 2013,
  \bibinfo{title}{emcee: The MCMC Hammer,} PASP, 125, 306,
  \dodoi{10.1086/670067}

\bibitem[{J.~M. Fregeau {et~al.}(2006)Fregeau, Chatterjee, \&
  Rasio}]{Fregeau_2006}
Fregeau, J.~M., Chatterjee, S., \& Rasio, F.~A. 2006, \bibinfo{title}{Dynamical
  Interactions of Planetary Systems in Dense Stellar Environments,} The
  Astrophysical Journal, 640, 1086, \dodoi{10.1086/500111}

\bibitem[{J.~P. Gardner {et~al.}(2006)Gardner {et~al.}}]{gardner2006jwst}
Gardner, J.~P., {et~al.} 2006, \bibinfo{title}{The James Webb Space Telescope,}
  Space Science Reviews, 123, 485, \dodoi{10.1007/s11214-006-8315-7}

\bibitem[{C. Hellier {et~al.}(2017)Hellier {et~al.}}]{hellier2017wasp}
Hellier, C., {et~al.} 2017, \bibinfo{title}{The discovery of WASP-127b,
  WASP-136b, and WASP-138b,} Astronomy \& Astrophysics, 601, A53,
  \dodoi{10.1051/0004-6361/201629294}

\bibitem[{M.~J. Holman \& N.~W. Murray(2005)Holman \& Murray}]{holman2005ttv}
Holman, M.~J., \& Murray, N.~W. 2005, \bibinfo{title}{The use of transit timing
  to detect terrestrial-mass extrasolar planets,} Science, 307, 1288,
  \dodoi{10.1126/science.1107822}

\bibitem[{J.~D. Hunter(2007)Hunter}]{Hunter:2007}
Hunter, J.~D. 2007, \bibinfo{title}{Matplotlib: A 2D graphics environment,}
  Computing in Science \& Engineering, 9, 90, \dodoi{10.1109/MCSE.2007.55}

\bibitem[{T. Kluyver {et~al.}(2016)Kluyver, Ragan-Kelley, P{\'e}rez, Granger,
  Bussonnier, Frederic, Kelley, Hamrick, Grout, Corlay, Ivanov, Avila, Abdalla,
  Willing, \& development team}]{jupyter}
Kluyver, T., Ragan-Kelley, B., P{\'e}rez, F., {et~al.} 2016, in Positioning and
  Power in Academic Publishing: Players, Agents and Agendas, ed. F.~Loizides \&
  B.~Scmidt (Netherlands: IOS Press), 87--90.
\newblock \url{https://eprints.soton.ac.uk/403913/}

\bibitem[{A. Lamberts {et~al.}(2019)Lamberts, Blunt, Littenberg,
  Garrison-Kimmel, Kupfer, \& Sanderson}]{lamberts2019predicting}
Lamberts, A., Blunt, S., Littenberg, T.~B., {et~al.} 2019,
  \bibinfo{title}{Predicting the LISA white dwarf binary population in the
  Milky Way with cosmological simulations,} Monthly Notices of the Royal
  Astronomical Society, 490, 5888, \dodoi{10.1093/mnras/stz2834}

\bibitem[{R.~A. {Mardling} \& S.~J. {Aarseth}(2001){Mardling} \&
  {Aarseth}}]{2001MNRAS.321..398M}
{Mardling}, R.~A., \& {Aarseth}, S.~J. 2001, \bibinfo{title}{{Tidal
  interactions in star cluster simulations},} \mnras, 321, 398,
  \dodoi{10.1046/j.1365-8711.2001.03974.x}

\bibitem[{C. Marois {et~al.}(2008)Marois {et~al.}}]{marois2008direct}
Marois, C., {et~al.} 2008, \bibinfo{title}{Direct imaging of multiple planets
  orbiting the star HR 8799,} Science, 322, 1348,
  \dodoi{10.1126/science.1166585}

\bibitem[{M. Mayor \& D. Queloz(1995)Mayor \& Queloz}]{mayor1995jupiter}
Mayor, M., \& Queloz, D. 1995, \bibinfo{title}{A Jupiter-mass companion to a
  solar-type star,} Nature, 378, 355, \dodoi{10.1038/378355a0}

\bibitem[{M. Mayor {et~al.}(2003)Mayor {et~al.}}]{mayor2003harps}
Mayor, M., {et~al.} 2003, \bibinfo{title}{Setting New Standards with HARPS,}
  The Messenger, 114, 20

\bibitem[{T.~D. {Morton} {et~al.}(2016){Morton}, {Bryson}, {Coughlin}, {Rowe},
  {Ravichandran}, {Petigura}, {Haas}, \& {Batalha}}]{2016ApJ...822...86M}
{Morton}, T.~D., {Bryson}, S.~T., {Coughlin}, J.~L., {et~al.} 2016,
  \bibinfo{title}{{False Positive Probabilities for all Kepler Objects of
  Interest: 1284 Newly Validated Planets and 428 Likely False Positives},}
  \apj, 822, 86, \dodoi{10.3847/0004-637X/822/2/86}

\bibitem[{ {NASA Exoplanet Archive}(2020){NASA Exoplanet Archive}}]{ps}
{NASA Exoplanet Archive}. 2020, \bibinfo{title}{Planetary Systems,}
  NExScI-Caltech/IPAC, \dodoi{10.26133/NEA12}

\bibitem[{G.~R. Ricker {et~al.}(2015)Ricker {et~al.}}]{ricker2015tess}
Ricker, G.~R., {et~al.} 2015, \bibinfo{title}{Transiting Exoplanet Survey
  Satellite (TESS),} Journal of Astronomical Telescopes, Instruments, and
  Systems, 1, 014003, \dodoi{10.1117/1.JATIS.1.1.014003}

\bibitem[{T. Robson {et~al.}(2019)Robson, Cornish, \& Liu}]{Robson_2019}
Robson, T., Cornish, N.~J., \& Liu, C. 2019, \bibinfo{title}{The construction
  and use of LISA sensitivity curves,} Classical and Quantum Gravity, 36,
  105011, \dodoi{10.1088/1361-6382/ab1101}

\bibitem[{J. {Sahlmann} {et~al.}(2016){Sahlmann}, {Lazorenko}, {S{\'e}gransan},
  {Astudillo-Defru}, {Bonfils}, {Delfosse}, {Forveille}, {Hagelberg}, {Lo
  Curto}, {Pepe}, {Queloz}, {Udry}, \& {Zimmerman}}]{2016AandA...595A..77S}
{Sahlmann}, J., {Lazorenko}, P.~F., {S{\'e}gransan}, D., {et~al.} 2016,
  \bibinfo{title}{{The mass of planet GJ 676A b from ground-based astrometry. A
  planetary system with two mature gas giants suitable for direct imaging},}
  \aap, 595, A77, \dodoi{10.1051/0004-6361/201628854}

\bibitem[{K.~C. {Sahu} {et~al.}(2006){Sahu}, {Casertano}, {Bond}, {Valenti},
  {Ed Smith}, {Minniti}, {Zoccali}, {Livio}, {Panagia}, {Piskunov}, {Brown},
  {Brown}, {Renzini}, {Rich}, {Clarkson}, \& {Lubow}}]{2006Natur.443..534S}
{Sahu}, K.~C., {Casertano}, S., {Bond}, H.~E., {et~al.} 2006,
  \bibinfo{title}{{Transiting extrasolar planetary candidates in the Galactic
  bulge},} \nat, 443, 534, \dodoi{10.1038/nature05158}

\bibitem[{S. {Seager}(2010){Seager}}]{Exoplanets_s_seager_book_2010}
{Seager}, S. 2010, {Exoplanets}

\bibitem[{N. Seto(2008)Seto}]{seto2008detecting}
Seto, N. 2008, \bibinfo{title}{Detecting planets around compact binaries with
  gravitational wave detectors in space,} The Astrophysical Journal, 677, L55

\bibitem[{S. {Sigurdsson} {et~al.}(2003){Sigurdsson}, {Richer}, {Hansen},
  {Stairs}, \& {Thorsett}}]{2003Sci...301..193S}
{Sigurdsson}, S., {Richer}, H.~B., {Hansen}, B.~M., {Stairs}, I.~H., \&
  {Thorsett}, S.~E. 2003, \bibinfo{title}{{A Young White Dwarf Companion to
  Pulsar B1620-26: Evidence for Early Planet Formation},} Science, 301, 193,
  \dodoi{10.1126/science.1086326}

\bibitem[{N. Tamanini \& C. Danielski(2018)Tamanini \&
  Danielski}]{tamanini2018listening}
Tamanini, N., \& Danielski, C. 2018, \bibinfo{title}{Listening to the
  gravitational wave sound of circumbinary exoplanets,} arXiv preprint
  arXiv:1812.04330

\bibitem[{N. Tamanini \& C. Danielski(2019)Tamanini \&
  Danielski}]{tamanini2019gravitational}
Tamanini, N., \& Danielski, C. 2019, \bibinfo{title}{The gravitational-wave
  detection of exoplanets orbiting white dwarf binaries using LISA,} Nature
  Astronomy, 3, 858

\bibitem[{S.~E. Thorsett {et~al.}(1999)Thorsett, Arzoumanian, \&
  Taylor}]{thorsett1999pulsar}
Thorsett, S.~E., Arzoumanian, Z., \& Taylor, J.~H. 1999, \bibinfo{title}{The
  triple system PSR B1620−26 in M4,} The Astrophysical Journal, 523, 763,
  \dodoi{10.1086/307759}

\bibitem[{A. Tiwari {et~al.}(2026)Tiwari, Kapadia, Vijaykumar, \&
  Chatterjee}]{Tiwari:losv2026}
Tiwari, A., Kapadia, S.~J., Vijaykumar, A., \& Chatterjee, S. 2026,
  \bibinfo{title}{{Periodic line-of-sight velocity-driven modulations to
  gravitational waves emitted by compact binaries in Keplerian outer orbits},}
  \doarXiv{2607.09644}

\bibitem[{A. Tiwari {et~al.}(2025)Tiwari, Vijaykumar, Kapadia, Chatterjee, \&
  Fragione}]{Tiwari:2024pvb}
Tiwari, A., Vijaykumar, A., Kapadia, S.~J., Chatterjee, S., \& Fragione, G.
  2025, \bibinfo{title}{{Profiling stellar environments of gravitational wave
  sources},} Phys. Rev. D, 112, 084034, \dodoi{10.1103/gspl-m478}

\bibitem[{S. van~der Walt {et~al.}(2011)van~der Walt, Colbert, \&
  Varoquaux}]{vanderWalt:2011bqk}
van~der Walt, S., Colbert, S.~C., \& Varoquaux, G. 2011, \bibinfo{title}{{The
  NumPy Array: A Structure for Efficient Numerical Computation},} Comput. Sci.
  Eng., 13, 22, \dodoi{10.1109/MCSE.2011.37}

\bibitem[{P. Virtanen {et~al.}(2020)Virtanen {et~al.}}]{Virtanen:2019joe}
Virtanen, P., {et~al.} 2020, \bibinfo{title}{{SciPy 1.0--Fundamental Algorithms
  for Scientific Computing in Python},} Nature Meth.,
  \dodoi{10.1038/s41592-019-0686-2}

\bibitem[{P. {Vynatheya} {et~al.}(2022){Vynatheya}, {Hamers}, {Mardling}, \&
  {Bellinger}}]{2022MNRAS.516.4146V}
{Vynatheya}, P., {Hamers}, A.~S., {Mardling}, R.~A., \& {Bellinger}, E.~P.
  2022, \bibinfo{title}{{Algebraic and machine learning approach to
  hierarchical triple-star stability},} \mnras, 516, 4146,
  \dodoi{10.1093/mnras/stac2540}

\bibitem[{A. Wolszczan \& D. Frail(1992)Wolszczan \&
  Frail}]{wolszczan1992pulsar}
Wolszczan, A., \& Frail, D. 1992, \bibinfo{title}{A planetary system around the
  millisecond pulsar PSR1257+12,} Nature, 355, 145, \dodoi{10.1038/355145a0}

\end{thebibliography}

\appendix 

\section{Jacobian}\label{app: jac}

The Jacobian $\boldsymbol{J} \equiv \partial(\mathcal{M}, \eta, z_{\rm L,0}, \Omega_{\rm det}) / \partial(\mathcal{M}, \eta, M_{\rm pl}, a)$ of the transformation from $(\mathcal{M}, \eta, M_{\rm pl}, a )$ to $(\mathcal{M}, \eta, z_{\rm L,0}, \Omega_{\rm det})$ is given by

\begin{equation}
    \label{eq: jac}
    \boldsymbol{J} = \left(
        \begin{array}{cccc}
            1 & 0 & 0 & 0 \\
            0 & 1 & 0 & 0 \\
            -\frac{z_{\rm L,0}}{2(1 + z_{\rm cos}) \eta^{3/5}}\frac{M_{\odot}}{M_{\rm pl} + M_s } & \frac{3 z_{\rm L,0}}{10 \eta}\frac{M_s}{M_{\rm pl} + M_s }  & \frac{z_{\rm L,0}}{2} \frac{M_{\rm pl} + 2 M_s}{M_{\rm pl}  + M_s } \frac{M_{\rm J}}{M_{\rm pl}} & - \frac{z_{\rm L,0}}{2} \frac{\rm AU}{a} \\
            \frac{\Omega_{\rm det}}{2(1 + z_{\rm cos}) \eta^{3/5}}\frac{M_{\odot}}{M_{\rm pl} + M_s} & - \frac{3 \Omega_{\rm det}}{10 \eta} \frac{M_s}{M_{\rm pl} + M_s} & \frac{\Omega_{\rm det}}{2} \frac{M_{\rm J}}{M_{\rm pl} + M_s} \frac{a}{\rm AU} & - \frac{3\Omega_{\rm det}}{2} \frac{\rm AU}{a} \\
        \end{array}
        \right)
\end{equation}
The determinant of this Jacobian is given by
\begin{equation}
    \label{eq: det_J}
    \vert \boldsymbol{J} \vert = -  \frac{\rm AU}{a} \frac{M_{\rm J}}{M_{\rm pl}} \frac{M_{\rm pl} + 3 M_s}{2 (M_{\rm pl} + M_s)} z_{\rm L,0} \Omega_{\rm det}
\end{equation}
Therefore, the Jacobian $\boldsymbol{J}^{-1} \equiv \partial(\mathcal{M}, \eta, M_{\rm pl}, a) / \partial(\mathcal{M}, \eta, z_{\rm L,0}, \Omega_{\rm det})$ of the inverse transformation becomes
\begin{equation}
    \label{eq: jac_inv}
    \boldsymbol{J}^{-1} = \left(
        \begin{array}{cccc}
            1 & 0 & 0 & 0 \\
            0 & 1 & 0 & 0 \\
            \frac{2}{(1 + z_{\rm cos}) \eta^{3/5}}\frac{M_{\odot}}{M_{\rm pl} + 3 M_s } \frac{M_{\rm pl}}{M_{\rm J}} & - \frac{6}{5 \eta} \frac{M_s}{M_{\rm pl} + 3 M_s } \frac{M_{\rm pl}}{M_{\rm J}}  & \frac{3}{z_{\rm L,0}} \frac{M_{\rm pl} + M_s}{M_{\rm pl} + 3 M_s } \frac{M_{\rm pl}}{M_{\rm J}}  & - \frac{1}{\Omega_{\rm det}} \frac{M_{\rm pl} + M_s}{M_{\rm pl} + 3 M_s } \frac{M_{\rm pl}}{M_{\rm J}} \\
            \frac{1}{(1 + z_{\rm cos}) \eta^{3/5}}\frac{M_{\odot}}{M_{\rm pl} + 3 M_s} \frac{a}{\rm AU} & - \frac{3}{5 \eta} \frac{M_s}{M_{\rm pl} + 3 M_s} \frac{a}{\rm AU} & \frac{1}{z_{\rm L,0}} \frac{M_{\rm pl}}{M_{\rm pl} + 3 M_s} \frac{a}{\rm AU} & - \frac{1}{\Omega_{\rm det}} \frac{M_{\rm pl} + 2 M_s}{M_{\rm pl} + 3 M_s} \frac{a}{\rm AU} \\
        \end{array}
        \right)
\end{equation}

\section{Priors}\label{app: priors}
Table~\ref{tab: priors} shows the priors used in the analysis during sampling for eccentric outer orbit examples presented in Figure~\ref{fig: 1d_posts}. 

\begin{table}[ht!]
    \centering
    \begin{tabular}{|l|r|}
         \hline
         \textbf{Parameter} & \textbf{Prior} \\
         \hline
         \hline
         Detector frame chirp mass $\mathcal{M} \, [\rm \mathcal{M}]$ & $\mathcal{U}(10^{-1}, \, 50)$ \\
         Symmetric mass ratio $\eta$ & $\mathcal{U}(10^{-3}, \, 0.25)$ \\
         Mass of the exoplanet $M_{\rm pl} \, [\rm M_J]$ & $\mathcal{U}(10^{-5}, \, 150)$ \\
         Semi-major axis of the exoplanet's orbit around the CBC $a \, [\rm AU]$ & $\mathcal{U}(10^{-5}, \, 100)$ \\
         Eccentricity of the outer orbit $e_{\rm out}$ & $\mathcal{U}(0, \, 1)$ \\
         \hline
    \end{tabular}
    \caption{Table of priors used during sampling of the parameters for eccentric outer orbit scenarios considered in Figure~\ref{fig: 1d_posts}. $\mathcal{U}$ indicates the uniform prior.}
    \label{tab: priors}
\end{table}

\section{Additional Figures}\label{app: add_fig}
\begin{figure}
    \centering
    \includegraphics[width=0.55\linewidth]{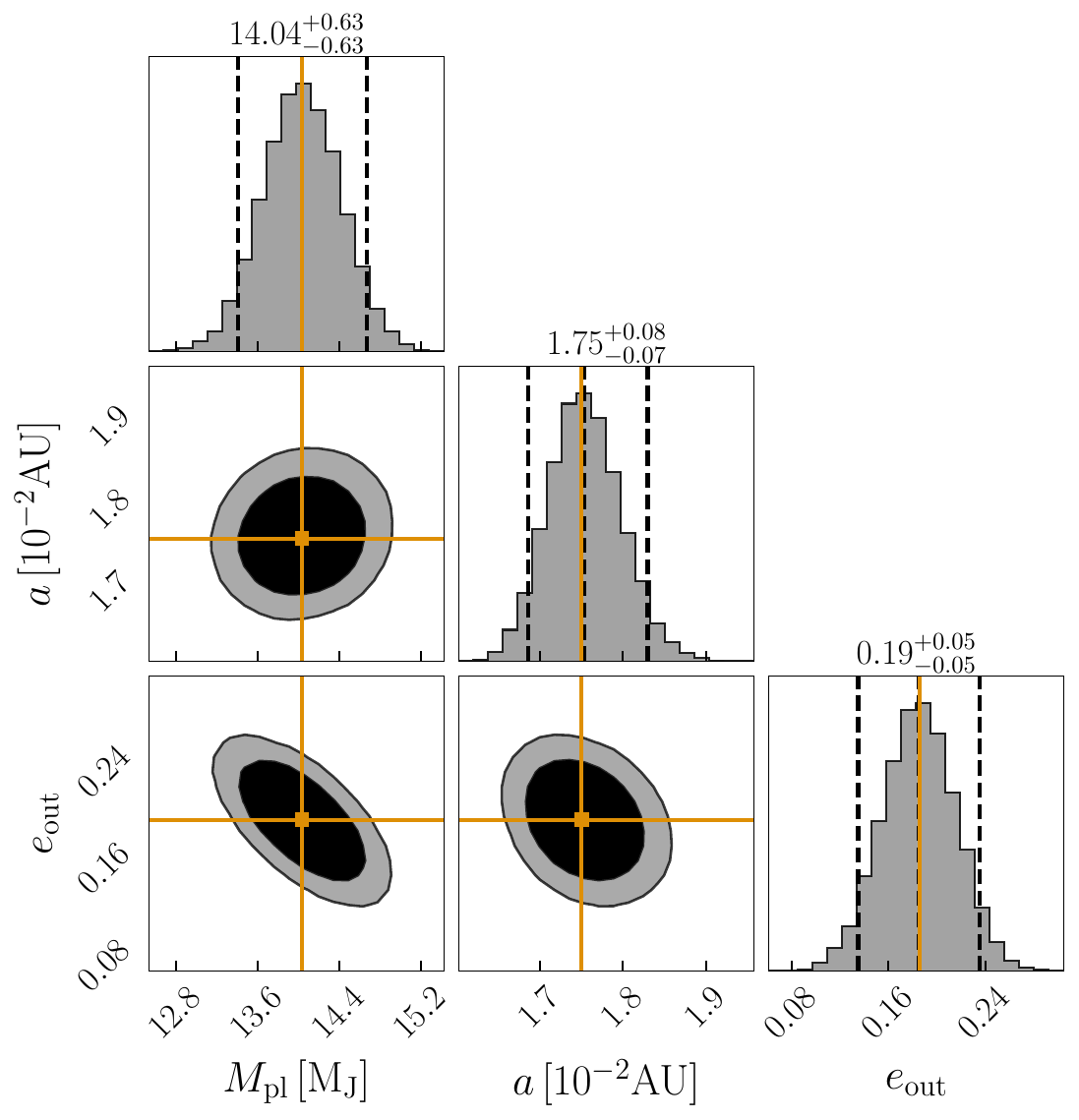}
    \caption{\textbf{ET: BNS}: the corner plot of $M_{\rm pl}$, $a$, and $e_{\rm out}$ for the ET:BNS scenario presented in the Figure~\ref{fig: 1d_posts}. The solid lines represent the true values, while the dashed lines represent the 90\% credible intervals. The 2D contours in the off-diagonal subplots correspond to the 68\% and 90\% credible regions.}
    \label{fig: ET_BNS_corn_plot}
\end{figure}

\begin{figure}
    \centering
    \includegraphics[width=0.48\linewidth]{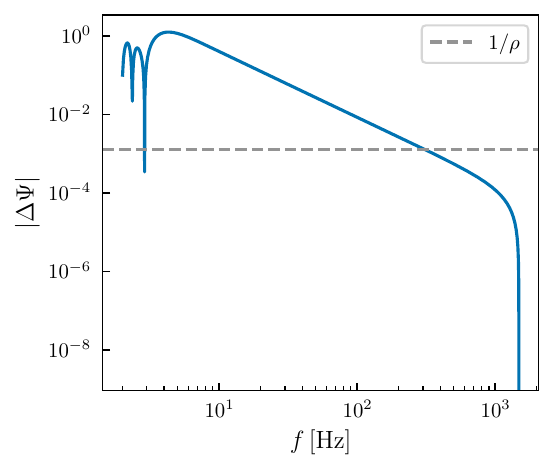}
    \includegraphics[width=0.45\linewidth]{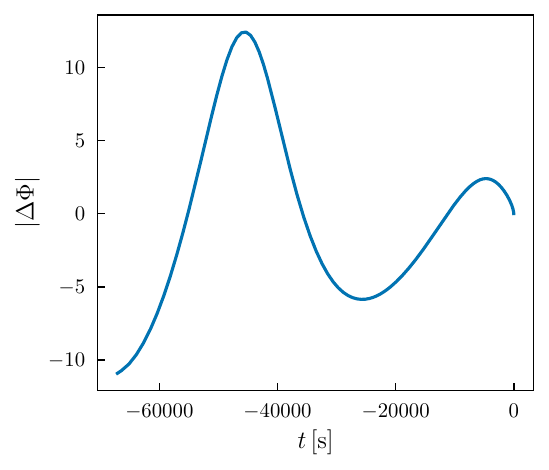}
    \caption{The frequency (\textit{left panel}) and time (\textit{right panel}) domain phase corrections for the ET: BNS scenario presented in Figure~\ref{fig: 1d_posts}. The dashed line in the left panel represents the minimum measurable phase correction $\sim 1/\rho$. The oscillations in $\Delta \Psi$ at lower frequencies are because at lower frequencies $\xi/v^8 \gg 1$ (see Equation~\ref{eq: ph_cor_losv_circ}), while as the frequency of GWs evolves, we reach the $\xi/v^8 \ll 1$ region and hence oscillations vanish.}
    \label{fig: phase_corr_et_bns}
\end{figure}

Figure~\ref{fig: ET_BNS_corn_plot} shows the corner plot of $M_{\rm pl}$, $a$, and $e_{\rm out}$ after placing \textit{Kepler-80 f} \citep{2016ApJ...822...86M} in an eccentric orbit around a 1.6 - 1.3 $M_{\odot}$ BNS at 100 Mpc merging in the ET band. Notice that the true values of all parameters are recovered within the 90\% credible level. In Figure~\ref{fig: phase_corr_et_bns}, we also show the corresponding frequency and time domain phase corrections, where time and time domain phase corrections have been computed using $t = (1/2\pi) d\Psi / df$ and $\Delta \Phi = f (d\Delta \Psi / df) - \Delta \Psi$, respectively. Here $\Psi$ and $\Delta \Psi$ are the total phase and phase correction in the frequency domain, respectively.

\end{document}